\documentclass[conference]{IEEEtran}
\IEEEoverridecommandlockouts

\usepackage{cite}
\usepackage{amsmath,amssymb,amsfonts,amsthm,mathtools,bm}
\usepackage{graphicx}
\usepackage{textcomp}
\usepackage{xcolor}
\usepackage{booktabs}
\usepackage{array}
\usepackage{enumitem}
\usepackage{microtype}
\usepackage{placeins}
\usepackage{url}
\usepackage{hyperref}
\hypersetup{colorlinks=true,linkcolor=blue,citecolor=blue,urlcolor=blue}
\newtheorem{theorem}{Theorem}
\newtheorem{proposition}{Proposition}
\newtheorem{corollary}{Corollary}

\newcommand{\X}{\mathcal X}
\newcommand{\Y}{\mathcal Y}
\newcommand{\Z}{\mathbb Z}

\newcommand{\pos}[1]{\left[#1\right]_+}
\newcommand{\RR}{\mathrm{RR}}

\begin{document}

\title{Exact Privacy--Distortion Design of\\
Support-Constrained Randomized Response}
%{\large Extended Version}}

\author{%
\begin{tabular}{ccc}
\begin{minipage}[t]{0.31\textwidth}
\centering 1\textsuperscript{st}
Amirreza Zamani\\
\textit{Department of Information Science and Engineering}\\
\textit{KTH Royal Institute of Technology}\\
Stockholm, Sweden\\
amizam@kth.se
\end{minipage}
&
\begin{minipage}[t]{0.31\textwidth}
\centering 2\textsuperscript{nd}
Parastoo Sadeghi\\
\textit{School of Engineering and Technology}\\
\textit{UNSW Canberra}\\
Canberra, Australia\\
p.sadeghi@unsw.edu.au
\end{minipage}
&
\begin{minipage}[t]{0.31\textwidth}
\centering 3\textsuperscript{rd}
Mikael Skoglund\\
\textit{Department of Information Science and Engineering}\\
\textit{KTH Royal Institute of Technology}\\
Stockholm, Sweden\\
skoglund@kth.se
\end{minipage}
\end{tabular}}

\maketitle

\begin{abstract}
Full-support local randomizers can incur unnecessary reporting distortion on ordered alphabets. We study randomized response whose output is restricted to a radius-$t$ neighborhood of the private input. Input-dependent supports create boundary leakage, so ordinary likelihood-ratio calibration is insufficient. Under approximate local privacy restricted to input pairs at distance at most $H$, we derive the exact pairwise and range-$H$ hockey-stick defects. We solve the joint design of the truthful probability, off-truth probability, and support radius in closed form at every radius, and reduce global optimization to at most $\lfloor H/2\rfloor+1$ candidates. We characterize the exact privacy floor, its inverse frontier under a distortion budget, the complete interval of feasible probabilities, and the support-size requirement. In the stringent-privacy limit, the optimal support grows as $H/\delta$ and the minimum line distortion grows as $H/(4\delta)$. The calibration extends unchanged to arbitrary nonnegative symmetric displacement costs at a fixed radius. Deterministic evaluations verify the formulas and expose the roles of the privacy range and approximation budget.

\end{abstract}


\section{Introduction}

In the local model of differential privacy, each user randomizes a sample before it is collected. The server therefore observes only privatized reports and does not need to be trusted with raw data. Classical randomized response and its $k$-ary variants assign positive probability to every report~\cite{warner1965,duchi2018,kairouz2016}. This common-support structure is natural for all-pairs local differential privacy, but it ignores geometry: confusing a rating of one with a rating of two can be much less damaging than confusing it with the largest rating. The same issue arises for integer counts, quantized measurements, location bins, and other ordered symbols.

We consider a support-constrained randomizer that reports only within a radius-$t$ neighborhood of its input. The radius is a design variable rather than a fixed implementation choice. A smaller radius suppresses long reports, but it also makes the input-dependent supports easier to distinguish. In particular, observing a report that belongs to one neighborhood but not another rules out the second input. Approximate privacy can absorb this mismatch through its $\delta$ budget, provided that support leakage and likelihood-ratio excess on the common support are accounted for jointly. This creates a genuine three-way design problem over the support radius, truthful mass, and off-truth mass.

Metric and policy-based privacy notions restrict which input pairs must be difficult to distinguish~\cite{andres2013,he2014}. We use a simple relational guarantee that protects pairs separated by at most a prescribed range $H$. This guarantee is intentionally weaker than ordinary all-pairs local privacy on an unbounded alphabet; the protected relation is part of the public privacy specification. Sparse distribution estimation and optimized local channels address complementary statistical questions~\cite{acharya2021,kairouz2014,murakami2019}. Recent work has analyzed sparse Laplace and Gaussian kernels~\cite{zamani2026sparse}. Our focus is different: we derive an exact mechanism-level theory for support-constrained randomized response, including the full feasible set and globally optimal line-distortion design.

The first contribution is an exact privacy account. We decompose the hockey-stick defect into support leakage and overlap excess, prove the common-support obstruction under pure privacy, and evaluate the defect for every separation. Maximization over the range relation then yields a closed-form scalar constraint.

The second contribution is an exact design solution. For every radius, we characterize the complete interval of privacy-feasible probabilities and obtain the unique line-distortion optimizer. We identify the exact privacy floor, derive its inverse as the best defect under a distortion budget, and show that global optimization needs at most $\lfloor H/2\rfloor+1$ radius evaluations rather than an unbounded search.

The third contribution is a collection of analytical consequences. Stringent privacy forces support size at least $1+H/\delta-e^\varepsilon$. As $\delta$ vanishes, the globally optimal support and distortion obey $s^*=H/\delta+O(1)$ and $D^*=H/(4\delta)+O(1)$. We quantify exactly when approximate calibration improves on the usual likelihood-ratio calibration and extend the fixed-radius solution to arbitrary nonnegative symmetric displacement costs.

The experiments are deterministic evaluations of the derived expressions. No privacy quantity is inferred by Monte Carlo simulation. %We intentionally exclude distribution-estimation risk, cyclic spectral analysis, minimax robustness, nonhomogeneous path mechanisms, distribution shift, and adaptive channel selection. Those results belong to the longer journal study and are not needed for the focused workshop contribution developed here.

The remainder of the paper introduces the relational privacy model, derives the exact defect, solves the joint design, studies analytical consequences, and presents numerical evaluations. The appendix contains complete proofs without repeating the result statements.

\section{Related Work and Positioning}

Randomized response originated as a survey method for eliciting sensitive binary attributes~\cite{warner1965}. Modern local differential privacy generalizes that idea to arbitrary channels and statistical tasks. Minimax analyses clarify the price of local privacy in estimation~\cite{duchi2018}, while distribution-estimation work studies channel choice and decoding on larger alphabets~\cite{kairouz2016,murakami2019}. Extremal-mechanism results further show that useful local channels often have highly structured probability matrices~\cite{kairouz2014}. These works largely begin with a common output support, which is unavoidable under finite pure all-pairs privacy.

Ordered and spatial data motivate privacy specifications that incorporate a public notion of adjacency or distance. Geo-indistinguishability scales privacy loss with geographic distance~\cite{andres2013}; policy-based formulations such as Blowfish restrict the pairs whose distinction must be hidden~\cite{he2014}. Our range-$H$ relation is a deliberately simple policy on the integer line. Unlike metric privacy, it places one common $(\varepsilon,\delta)$ bound on every pair within the declared range. Unlike ordinary local privacy, it makes no promise for pairs farther apart. This distinction is essential because any finite-radius channel on an unbounded alphabet has disjoint supports at sufficiently large separations.

Two uses of the word sparse should also be separated. Sparse distribution estimation assumes that the unknown population law has few nonzero coordinates and studies the statistical complexity of learning it~\cite{acharya2021}. Mechanism sparsity instead means that a channel row assigns zero probability to many reports. The present work concerns the latter. Restricting a channel row can reduce displacement cost, yet the resulting zeros create an explicit privacy charge. Sparse Laplace and Gaussian kernels provide related constructions for ordered data~\cite{zamani2026sparse}; randomized response is sufficiently structured to permit a complete closed-form analysis.

The hockey-stick expression used below is the discrete $E_\gamma$ divergence with $\gamma=e^\varepsilon$. It is equivalent to the eventwise formulation of approximate differential privacy~\cite{dwork2014} and connects local privacy to divergence contraction~\cite{asoodeh2021}. Its positive-part form is especially convenient for support-constrained channels because it separates outputs missing from the competing row from outputs present in both rows. Our contribution is to carry that accounting through an exact joint mechanism design, rather than to introduce a new privacy definition.

\section{Relational Local Privacy and Support Leakage}

Let $M(\cdot\mid x)$ be a channel from a discrete input alphabet $\X$ to a discrete output alphabet $\Y$. For a symmetric relation $\mathcal R\subseteq\X\times\X$, the channel is $(\varepsilon,\delta)$-locally differentially private on $\mathcal R$ if
\begin{equation}
 \Delta_\varepsilon^M(x,x')
 :=\sum_{y\in\Y}\pos{M(y\mid x)-e^\varepsilon M(y\mid x')}
 \leq \delta
 \label{eq:hockey}
\end{equation}
for every $(x,x')\in\mathcal R$. Equation~\eqref{eq:hockey} is the discrete hockey-stick representation of approximate local privacy~\cite{dwork2014,asoodeh2021}. It becomes ordinary all-pairs local differential privacy when $\mathcal R=\X\times\X$.

Our ordered model uses $\X=\Y=\Z$ and protects the range-$H$ relation
\begin{equation}
 \mathcal R_H=\{(x,x'):1\leq |x-x'|\leq H\}.
 \label{eq:relation}
\end{equation}
This is weaker than all-pairs privacy on $\Z$; its interpretation and the protected range must therefore be reported explicitly.

Let $S(x)=\operatorname{supp}M(\cdot\mid x)$. Splitting the sum in~\eqref{eq:hockey} into outputs that disappear and outputs that remain possible gives
\begin{align}
\Delta_\varepsilon^M(x,x')
={}&M(S(x)\setminus S(x')\mid x) \nonumber\\
&+\sum_{y\in S(x)\cap S(x')}
\pos{M(y\mid x)-e^\varepsilon M(y\mid x')}.
\label{eq:decomposition}
\end{align}
The first term is \emph{support leakage}; the second is \emph{overlap excess}. This identity also explains the obstruction under pure privacy. Pure privacy cannot tolerate an output that is possible under one protected input and impossible under the other. The following statement isolates the exact common-support requirement.

\begin{proposition}
If $M$ satisfies pure $\varepsilon$-local differential privacy on a symmetric relation $\mathcal R$ for finite $\varepsilon$, then $S(x)=S(x')$ for every $(x,x')\in\mathcal R$. In particular, no finite-radius channel on $\Z$ satisfies pure all-pairs local differential privacy.
\label{prop:pure}
\end{proposition}

The proof is given in Appendix~\ref{proof:pure}.

We therefore treat $\delta$ as an explicit budget for boundary leakage and protect only the declared relation~\eqref{eq:relation}. Equality of $\X$ and $\Y$ is a modeling choice: each private integer is reported in the same ordered domain. It is not required by local differential privacy, which allows different input and output alphabets. Moreover, limiting the support does not by itself reduce communication; a local offset can be encoded with fewer symbols only when the decoder has a public anchor identifying the relevant neighborhood.

\section{Exact Defect of Sparse Randomized Response}

For an integer radius $t\geq1$, define
$S_t(x)=\{x-t,\ldots,x+t\}$ and $s=2t+1$. Sparse randomized response is
\begin{equation}
M^{\RR}(y\mid x)=
\begin{cases}
p, & y=x,\\
q, & y\in S_t(x)\setminus\{x\},\\
0, & y\notin S_t(x),
\end{cases}
\label{eq:sparse-rr}
\end{equation}
where $p\geq q\geq0$ and $p+2tq=1$. The truthful-bias constraint $p\geq q$ is equivalent to $0\leq q\leq 1/(2t+1)$.

The privacy defect depends only on the separation $h=|x-x'|$. Three geometries occur. The supports are disjoint when $h>2t$. When $h\leq t$, their mismatch contains only off-truth reports, while the truthful report remains in the overlap. When $t<h\leq2t$, the truthful report itself lies in the mismatch. Evaluating~\eqref{eq:decomposition} in these regimes gives the following exact expression.

\begin{proposition}
\label{prop:pair-defect}
For inputs separated by $h=|x-x'|$, the ordered defect of~\eqref{eq:sparse-rr} is
\begin{equation}
\Delta^{\RR}_\varepsilon(h)=
\begin{cases}
1, & h>2t,\\
hq+\pos{p-e^\varepsilon q}, & 1\leq h\leq t,\\
p+q(h-1), & t<h\leq2t.
\end{cases}
\label{eq:pair-defect}
\end{equation}
\end{proposition}

The proof is given in Appendix~\ref{proof:pair-defect}.

The defect in each overlapping regime is nondecreasing in $h$. Maximizing over~\eqref{eq:relation} therefore requires only the endpoint $H$ and produces a scalar privacy constraint.

\begin{corollary}
\label{cor:range-defect}
The exact range-$H$ defect is
\begin{equation}
\delta_t(p,q;\varepsilon,H)=
\begin{cases}
1, & H>2t,\\
Hq+\pos{p-e^\varepsilon q}, & H\leq t,\\
1-(2t-H+1)q, & t<H\leq2t.
\end{cases}
\label{eq:range-defect}
\end{equation}
\end{corollary}

The proof is given in Appendix~\ref{proof:range-defect}.

For the line-metric cost, the expected reporting distance is independent of the input and equals
\begin{equation}
D(t,p,q)=\sum_y |y-x|M^{\RR}(y\mid x)=qt(t+1).
\label{eq:distortion}
\end{equation}
Thus the off-truth probability controls distortion at a fixed radius, whereas the exact privacy constraint is piecewise linear in the same variable.

\section{Joint Privacy--Distortion Design}

Fix $\varepsilon>0$, $\delta\in(0,1)$, and $H\geq1$. After substituting $p=1-2tq$, the admissible interval is $0\leq q\leq1/(2t+1)$. Equation~\eqref{eq:distortion} increases in $q$, so the optimal fixed-radius design is the leftmost point at which~\eqref{eq:range-defect} reaches the target. The next theorem gives this point and the resulting global search.

\begin{theorem}
\label{thm:joint}
For a fixed radius $t$, the distortion-minimizing privacy-feasible probability is
\begin{equation}
q_t^*(\delta)=
\begin{cases}
\dfrac{1-\delta}{2t+e^\varepsilon-H},
& H\leq t,\quad \delta\geq\dfrac{H}{e^\varepsilon+2t},\\[1.1ex]
\dfrac{1-\delta}{2t-H+1},
& t<H\leq2t,\quad \delta\geq\dfrac{H}{2t+1},\\[1.1ex]
\text{infeasible}, & \text{otherwise},
\end{cases}
\label{eq:qstar}
\end{equation}
with $p_t^*(\delta)=1-2tq_t^*(\delta)$. The global optimum is
\begin{equation}
t^*\in\arg\min_{t:q_t^*(\delta)\ \mathrm{feasible}}
q_t^*(\delta)t(t+1).
\label{eq:tstar}
\end{equation}
The feasible set is nonempty, and the minimum in~\eqref{eq:tstar} is attained.
\end{theorem}

The proof is given in Appendix~\ref{proof:joint}.

The privacy constraint also admits a complete interval description. This is useful whenever reporting distance is replaced by another scalar channel criterion, because every fixed-radius design still reduces to a one-dimensional constrained problem.

\begin{proposition}
\label{prop:qinterval}
For every radius whose privacy floor does not exceed $\delta$, the feasible off-truth probabilities form the interval
\begin{equation}
\mathcal Q_t(\delta)=
\begin{cases}
\left[\dfrac{1-\delta}{2t+e^\varepsilon-H},
\min\!\left\{\dfrac{\delta}{H},\dfrac{1}{2t+1}\right\}\right],&H\leq t,\\[2ex]
\left[\dfrac{1-\delta}{2t-H+1},\dfrac{1}{2t+1}\right],&t<H\leq2t.
\end{cases}
\label{eq:qinterval}
\end{equation}
For all other radii, $\mathcal Q_t(\delta)$ is empty.
\end{proposition}

The proof is given in Appendix~\ref{proof:qinterval}.

Closed-form calibration also exposes the direction of every local design response. The statement below is conditional on staying inside one feasible overlap regime; crossing a feasibility boundary can change the optimal radius and is handled separately by the global candidate search.

\begin{proposition}
\label{prop:sensitivity}
At a fixed feasible radius, $q_t^*(\delta)$ and the optimized line distortion are strictly decreasing in $\delta$, while $p_t^*(\delta)$ is strictly increasing. Within either overlap regime, increasing $H$ strictly increases $q_t^*$ and distortion and strictly decreases $p_t^*$. In the strong-overlap regime $H\leq t$, increasing $\varepsilon$ strictly decreases $q_t^*$ and distortion; in the partial-overlap regime $t<H\leq2t$, the optimizer is independent of $\varepsilon$.
\end{proposition}

The proof is given in Appendix~\ref{proof:sensitivity}.

The support-leakage decomposition can also be evaluated at the optimum. This shows how the approximate budget is spent. In strong overlap, a target above the privacy floor is reached by reducing boundary mass and permitting a positive excess on the truthful output. In partial overlap, the truthful output is absent from the competing support, so the entire defect is unavoidable mismatch mass.

\begin{proposition}
\label{prop:budget}
Let $L_t^*$ and $E_t^*$ denote the support-leakage and overlap-excess terms in~\eqref{eq:decomposition} for a worst protected pair under the fixed-radius optimizer. If $H\leq t$ and $a=2t+e^\varepsilon$, then
\begin{equation}
L_t^*=\frac{H(1-\delta)}{a-H},\qquad
E_t^*=\frac{a\delta-H}{a-H},\qquad
L_t^*+E_t^*=\delta.
\label{eq:budget-split}
\end{equation}
If $t<H\leq2t$, then $L_t^*=\delta$ and $E_t^*=0$.
\end{proposition}

The proof is given in Appendix~\ref{proof:budget}.

For $(\varepsilon,H,\delta,t)=(1,3,0.45,3)$, the optimal channel assigns $0.28855$ of the defect to boundary leakage and $0.16145$ to truthful-output excess. At the privacy floor the second number is zero; all of the minimum attainable defect is boundary leakage. The decomposition therefore provides a direct diagnostic for whether a proposed calibration is limited by support geometry or by the likelihood ratio on the overlap.

The feasibility thresholds used in the theorem are themselves operationally useful. They quantify the smallest privacy defect possible at a prescribed support radius, independently of the reporting cost.

\begin{corollary}
\label{cor:floor}
The exact privacy floor at radius $t$ is
\begin{equation}
\delta_t^{\min}(\varepsilon,H)=
\begin{cases}
1, & H>2t,\\
\dfrac{H}{e^\varepsilon+2t}, & H\leq t,\\[0.8ex]
\dfrac{H}{2t+1}, & t<H\leq2t.
\end{cases}
\label{eq:floor}
\end{equation}
Hence a target $\delta$ is feasible at radius $t$ exactly when
$\delta\geq\delta_t^{\min}(\varepsilon,H)$.
\end{corollary}

The proof is given in Appendix~\ref{proof:floor}.

The preceding results fix a privacy target and minimize distortion. The dual design question fixes an allowable line distortion $B$ and asks for the strongest attainable privacy defect. Because the defect is V-shaped in the strong-overlap regime and decreasing in the partial-overlap regime, simply spending all of $B$ can be suboptimal once the privacy floor has been reached. The exact inverse frontier therefore saturates at that floor.

\begin{proposition}
\label{prop:inverse-budget}
For $B\geq0$, let $\underline\delta_t(B)$ be the minimum range-$H$ defect among radius-$t$ truthful-bias channels with $D(t,p,q)\leq B$. Then
\begin{equation}
\underline\delta_t(B)=
\begin{cases}
1, & H>2t,\\[0.4ex]
\displaystyle\max\!\left\{\frac{H}{2t+e^\varepsilon},
1-\frac{(2t+e^\varepsilon-H)B}{t(t+1)}\right\}, & H\leq t,\\[2.2ex]
\displaystyle\max\!\left\{\frac{H}{2t+1},
1-\frac{(2t-H+1)B}{t(t+1)}\right\}, & t<H\leq2t.
\end{cases}                                                        \label{eq:inverse-budget}
\end{equation}
The best defect across all radii is
$\underline\delta(B)=\min_{t\geq1}\underline\delta_t(B)$, and this minimum is attained.
\end{proposition}

The proof is given in Appendix~\ref{proof:inverse-budget}.

Equivalently, a target $\delta<1$ is attainable at radius $t$ under budget $B$ exactly when the radius is privacy feasible and
$B\geq q_t^*(\delta)t(t+1)$. Thus~\eqref{eq:qstar} and~\eqref{eq:inverse-budget} are the forward and inverse descriptions of the same finite-radius privacy--distortion frontier.

The floor can be inverted for every privacy target, not only in the stringent regime. It decreases as the radius grows, but its algebra changes when the truthful output first enters the overlap of every protected pair. The next formula makes this transition explicit and gives the exact smallest implementable support.

\begin{corollary}
\label{cor:minradius}
Let $t_{\mathrm{feas}}$ be the smallest radius that can attain a target $\delta\in(0,1)$. Then
\begin{equation}
t_{\mathrm{feas}}=
\begin{cases}
\displaystyle
\max\!\left\{\left\lceil\frac H2\right\rceil,
\left\lceil\frac{H/\delta-1}{2}\right\rceil\right\},
&\displaystyle \delta\geq\frac{H}{2H-1},\\[3ex]
\displaystyle
\max\!\left\{H,
\left\lceil\frac{H/\delta-e^\varepsilon}{2}\right\rceil\right\},
&\displaystyle \delta<\frac{H}{2H-1}.
\end{cases}
\label{eq:minradius-all}
\end{equation}
The exact minimum support size is $2t_{\mathrm{feas}}+1$.
\end{corollary}

The proof is given in Appendix~\ref{proof:minradius}.

The floor also produces a converse on sparsity. In the stringent approximate-privacy regime, every radius below the protected range is infeasible. Inverting the remaining branch yields the support requirement below.

\begin{corollary}
\label{cor:support}
If $\delta<H/(2H-1)$, every feasible radius satisfies $t\geq H$, and the support size $s=2t+1$ obeys
\begin{equation}
s\geq 1+\frac{H}{\delta}-e^\varepsilon.
\label{eq:support-bound}
\end{equation}
Consequently, fixed $(H,\varepsilon)$ and decreasing $\delta$ force
$s=\Omega(H/\delta)$.
\end{corollary}

The proof is given in Appendix~\ref{proof:support}.

The common likelihood-ratio calibration $p=e^\varepsilon q$ operates at the privacy floor when $H\leq t$. It is therefore unnecessarily noisy whenever the target $\delta$ is strictly above that floor: the design in~\eqref{eq:qstar} lowers $q$, increases the truthful mass, and uses the available approximate-privacy budget to reduce distortion.

The support converse is nearly exact after integer rounding. At $t\geq H$, feasibility is equivalent to $2t\geq H/\delta-e^\varepsilon$. Hence the smallest feasible radius in the stringent-privacy regime is
\begin{equation}
t_{\min}=\max\!\left\{H,\left\lceil\frac{H/\delta-e^\varepsilon}{2}\right\rceil\right\}.
\label{eq:tmin}
\end{equation}
Equation~\eqref{eq:tmin} strengthens the scaling statement by identifying the exact integer threshold for this mechanism.

The apparently unbounded search in~\eqref{eq:tstar} can in fact be reduced to a small finite set. Radii below $H/2$ are infeasible because a protected pair has disjoint supports. Above $H$, the optimized distortion is strictly increasing, so only the first feasible radius in that regime can be optimal. This leaves the transition radii and one large-radius candidate.

\begin{corollary}
\label{cor:candidates}
Define
\begin{equation}
\begin{aligned}
t_0&=\max\!\left\{H,\left\lceil\frac{H/\delta-e^\varepsilon}{2}\right\rceil\right\},\\
\mathcal T_H&=\left\{t\in\Z:\left\lceil\frac H2\right\rceil\leq t<H,\right.\\[-0.4ex]
&\hspace{4.2em}\left.\delta\geq\frac{H}{2t+1}\right\}\cup\{t_0\}.
\end{aligned}
\label{eq:candidates}
\end{equation}
Then the global minimization in~\eqref{eq:tstar} is equivalent to minimizing over $\mathcal T_H$, which contains at most $\lfloor H/2\rfloor+1$ radii.
\end{corollary}

The proof is given in Appendix~\ref{proof:candidates}.

The same privacy calibration applies to a broader class of reporting costs. Let a displacement cost be specified by nonnegative values $g(1),g(2),\ldots$ and set $c(x,y)=g(|x-y|)$ with $g(0)=0$. At a fixed radius, every nontruthful displacement $j$ and $-j$ has probability $q$, so the expected cost remains linear in the single design variable.

\begin{corollary}
\label{cor:cost}
For any nonnegative symmetric displacement cost with $\sum_{j=1}^t g(j)>0$, the expected cost is
\begin{equation}
C_g(t,p,q)=2q\sum_{j=1}^t g(j).
\label{eq:generalcost}
\end{equation}
Consequently, the fixed-radius privacy-optimal probabilities remain those in~\eqref{eq:qstar}; only the final comparison across radii changes. For the power cost $g(j)=j^r$, the optimized criterion is $2q_t^*(\delta)\sum_{j=1}^t j^r$.
\end{corollary}

The proof is given in Appendix~\ref{proof:cost}.

The usual likelihood-ratio calibration can now be compared with the approximate-privacy optimum at the same radius. In the strong-overlap regime, setting $p=e^\varepsilon q$ gives $q_{\mathrm{LR}}=(2t+e^\varepsilon)^{-1}$ and operates exactly at the privacy floor. Any slack between the target and that floor can be converted into a smaller off-truth probability.

\begin{corollary}
\label{cor:ratio}
Suppose $H\leq t$ and $\delta\geq H/(2t+e^\varepsilon)$. Relative to likelihood-ratio calibration at the same radius, the distortion ratio is
\begin{equation}
\frac{D(t,p_t^*,q_t^*)}{D(t,p_{\mathrm{LR}},q_{\mathrm{LR}})}
=\frac{(1-\delta)(2t+e^\varepsilon)}{2t+e^\varepsilon-H}\leq1.
\label{eq:ratio}
\end{equation}
The inequality is strict exactly when the target lies strictly above the privacy floor.
\end{corollary}

The proof is given in Appendix~\ref{proof:ratio}.

The exact integer threshold also determines the stringent-privacy asymptotics. Once $\delta<H/(2H-1)$, no transition radius below $H$ remains feasible, and the monotonicity established above makes the first feasible large-radius channel globally optimal. Expanding its closed-form parameters yields the following scaling law.

\begin{corollary}
\label{cor:asymptotics}
For fixed $H$ and $\varepsilon$, as $\delta\downarrow0$ the global line-distance optimum satisfies
\begin{align}
t^*&=\frac{H}{2\delta}+O(1), & s^*&=\frac{H}{\delta}+O(1),\nonumber\\
q^*&=\frac{\delta}{H}+O(\delta^2), &
p^*&=\frac{e^\varepsilon\delta}{H}+O(\delta^2),\nonumber\\[-0.5ex]
D^*&=\frac{H}{4\delta}+O(1).
\label{eq:asymptotics}
\end{align}
\end{corollary}

The proof is given in Appendix~\ref{proof:asymptotics}.

\section{Numerical Illustration}

\begin{figure*}[t]
\centering
\includegraphics[width=0.94\textwidth]{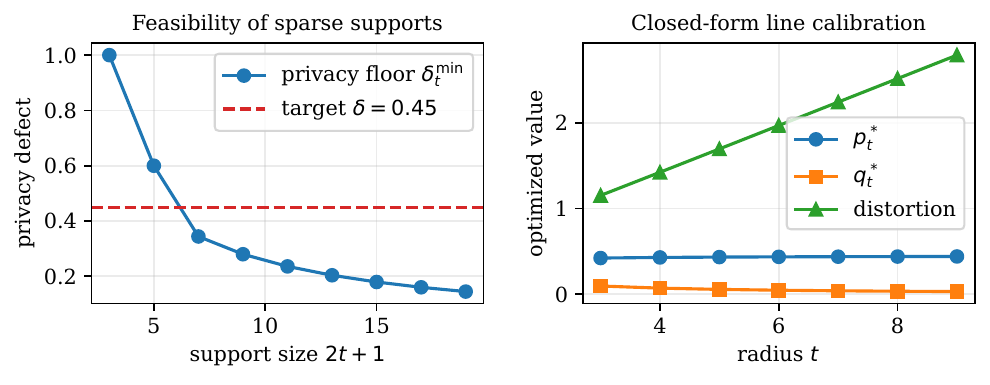}
\caption{Exact line-support design for $(\varepsilon,H,\delta)=(1,3,0.45)$. The left panel separates infeasible and feasible support sizes using~\eqref{eq:floor}. The right panel evaluates~\eqref{eq:qstar} and~\eqref{eq:distortion} on the feasible radii.}
\label{fig:design}
\end{figure*}

The formulas above require no stochastic simulation. Figure~\ref{fig:design} evaluates them for $(\varepsilon,H,\delta)=(1,3,0.45)$. The left panel plots the exact privacy floor against support size. The smallest feasible radius is $t=3$, corresponding to support size seven. The right panel reports the optimized probabilities and distortion for every feasible radius. Although larger supports permit smaller $q$, their longer reporting distances dominate; the global design is
\begin{equation}
(t^*,p^*,q^*)=(3,0.4229,0.09618),\qquad D^*=1.1542.
\end{equation}

\begin{figure}[t]
\centering
\includegraphics[width=0.98\columnwidth]{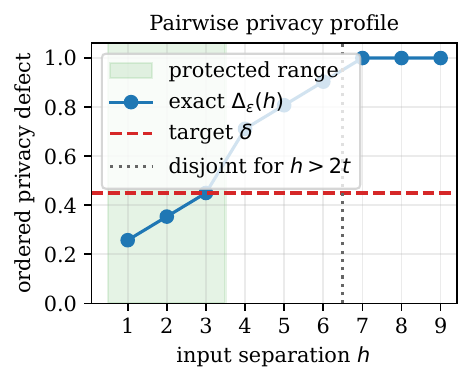}
\caption{Exact ordered defect versus input separation for the selected channel at $(\varepsilon,H,\delta)=(1,3,0.45)$. The shaded region is the protected relation.}
\label{fig:profile}
\end{figure}

The pairwise profile reveals what the range restriction does and does not protect. Figure~\ref{fig:profile} evaluates~\eqref{eq:pair-defect} for the selected channel. The defect increases from $0.2576$ at separation one to the target $0.45$ at the protected endpoint $h=3$. Beyond that endpoint, the same channel has larger defect; once $h>2t=6$, its supports are disjoint and the defect is one. Thus satisfying the declared range-$H$ guarantee must not be reported as satisfying ordinary all-pairs privacy.

At the same radius, the usual likelihood-ratio calibration uses
$q_{\mathrm{LR}}=(6+e)^{-1}=0.11470$ and has distortion $1.3764$. Its range-three defect is the privacy floor $0.3441$, so it leaves part of the permitted $\delta=0.45$ budget unused. Exact approximate calibration lowers the off-truth mass to $0.09618$ and the distortion to $1.1542$, a reduction of $16.15\%$. This numerical comparison is exactly the ratio in~\eqref{eq:ratio}, not a comparison between mechanisms with different advertised guarantees.

The support converse is most informative as $\delta$ decreases. For example, with $(\varepsilon,H)=(1,3)$ and $\delta=0.05$, the exact floor requires $t\geq29$, hence $s\geq59$, while~\eqref{eq:support-bound} gives $s\geq58.28$. The gap is due only to integer rounding. This agreement shows that support growth is not a proof artifact: it follows from the exact mechanism geometry.

Table~\ref{tab:delta} follows the globally distortion-optimal design as the privacy target tightens. Each row searches all integer radii using~\eqref{eq:tstar}; no truncation or Monte Carlo privacy test is used. The optimal radius equals the smallest feasible radius in these instances, while both support and distortion increase sharply as $\delta$ approaches zero.

\begin{table}[t]
\centering
\caption{Exact optimal designs for $\varepsilon=1$ and $H=3$.}
\label{tab:delta}
\setlength{\tabcolsep}{3.7pt}
\begin{tabular}{c|r|r|c|c}
$\delta$ & $t^*$ & $s^*$ & $q^*$ & $D^*$\\ \hline
$0.45$ & $3$  & $7$  & $0.09618$ & $1.154$\\
$0.30$ & $4$  & $9$  & $0.09069$ & $1.814$\\
$0.20$ & $7$  & $15$ & $0.05832$ & $3.266$\\
$0.10$ & $14$ & $29$ & $0.03247$ & $6.819$\\
$0.05$ & $29$ & $59$ & $0.01646$ & $14.320$
\end{tabular}
\end{table}

\begin{figure}[t]
\centering
\includegraphics[width=0.98\columnwidth]{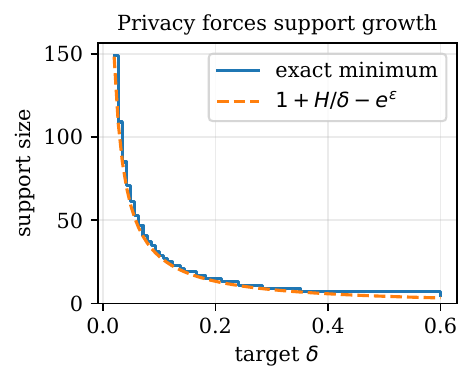}
\caption{Exact minimum feasible support size and analytical lower bound for $\varepsilon=1$ and $H=3$. The staircase is entirely due to the integer radius.}
\label{fig:scaling}
\end{figure}

The choice of $H$ is equally consequential. It defines the pairs protected by the relational guarantee. Increasing $H$ adds support-mismatch events to the privacy constraint and can force a larger radius and larger distortion. Thus $H$ should reflect a declared application-level indistinguishability range rather than being tuned on an individual's unprotected value.

\begin{figure}[t]
\centering
\includegraphics[width=0.96\columnwidth]{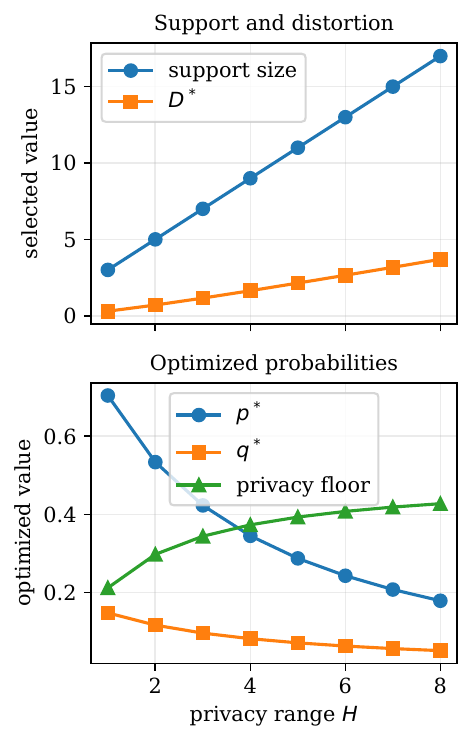}
\caption{Sensitivity of the exact distortion-optimal design to the declared privacy range at $(\varepsilon,\delta)=(1,0.45)$. A larger $H$ enlarges the protected relation and therefore requires wider supports and higher reporting distortion.}
\label{fig:range}
\end{figure}

Figure~\ref{fig:range} evaluates this effect at $(\varepsilon,\delta)=(1,0.45)$. The optimized support expands as $H$ grows, and the distortion rises because the mechanism must make more widely separated inputs difficult to distinguish. In all panels, privacy is evaluated from the exact ordered hockey-stick defect over every protected separation. The plotted probabilities are then obtained from~\eqref{eq:qstar}, so the computation has no privacy-estimation tolerance beyond floating-point evaluation of the closed forms.

Table~\ref{tab:range} reports representative points from the same sweep. For these parameters the minimizing radius equals $H$, but this is an output of the candidate comparison rather than an imposed constraint. The attained defect equals $0.45$ in every row; the last column is the smaller fixed-radius floor, which quantifies the amount of approximate-privacy slack converted into reduced distortion.

\begin{table*}[t]
\centering
\caption{Exact range sensitivity for $\varepsilon=1$ and $\delta=0.45$.}
\label{tab:range}
\setlength{\tabcolsep}{8pt}
\begin{tabular}{c|c|c|c|c|c|c}
$H$ & $t^*$ & $s^*$ & $p^*$ & $q^*$ & $D^*$ & $\delta_{t^*}^{\min}$\\ \hline
$1$ & $1$ & $3$  & $0.704164$ & $0.147918$ & $0.295836$ & $0.211942$\\
$2$ & $2$ & $5$  & $0.533729$ & $0.116568$ & $0.699407$ & $0.297695$\\
$3$ & $3$ & $7$  & $0.422904$ & $0.096183$ & $1.154193$ & $0.344104$\\
$4$ & $4$ & $9$  & $0.345071$ & $0.081866$ & $1.637323$ & $0.373194$\\
$6$ & $6$ & $13$ & $0.242970$ & $0.063086$ & $2.649605$ & $0.407656$\\
$8$ & $8$ & $17$ & $0.178973$ & $0.051314$ & $3.694622$ & $0.427390$
\end{tabular}
\end{table*}

\section{Design Implications and Scope}

The two overlap regimes have different operational meanings. When $t<H\leq2t$, a protected comparison places the truthful output outside the competing support. The defect is then determined entirely by mismatch mass and is independent of $\varepsilon$. Increasing the multiplicative privacy allowance cannot repair a zero probability in the competing row. In contrast, when $H\leq t$, every protected truthful output lies in the overlap; $\varepsilon$ then controls the likelihood-ratio term while $Hq$ accounts for boundary leakage. This distinction explains why expanding the radius to at least $H$ can be preferable even when a smaller radius is technically feasible.

Approximation is used here in a precise channel-level sense. The parameter $\delta$ bounds the largest ordered hockey-stick defect over the declared pairs; it is not estimated from samples and it is not a claim that every privacy violation occurs through one common input-independent failure event. Reporting the attained defect together with $H$ is therefore essential. A channel calibrated only for a smaller relation can exceed its advertised target when comparisons at larger separations are later included.

The idealized integer-line model removes edge effects and makes the support geometry transparent. On a bounded interval, naively truncating a neighborhood changes its normalization near the endpoints, so the formulas above should not be applied without recalculating the finite hockey-stick sum. A cyclic alphabet avoids endpoints but its supports can wrap around when the radius is large. Those geometries require their own exact accounting. They are deliberately excluded here, as are inverse distribution estimation, robust calibration under prior uncertainty, and context-dependent mechanism selection. This separation keeps the workshop contribution centered on the privacy--distortion law of line-supported randomized response.

From an implementation perspective, the formulas give a direct workflow. One first declares the public privacy range $H$ and the target $(\varepsilon,\delta)$. Equation~\eqref{eq:candidates} then produces a finite list of radii. For each candidate,~\eqref{eq:qstar} gives the unique distortion-minimizing probabilities and~\eqref{eq:range-defect} verifies the attained defect. The selected channel can therefore be audited without sampling, numerical optimization, or pairwise enumeration over the infinite alphabet.
\section{Conclusion}

Support-constrained randomized response admits exact privacy--distortion design on the integer line. Separating boundary leakage from overlap excess yields closed-form fixed-radius calibration, an attained global search, and the inverse distortion-budget frontier. The privacy floor forces support of order $H/\delta$ under strong approximate privacy, quantifying the geometric cost without asymptotic or Monte Carlo approximation.

\appendix
\section{Proofs}

\subsection{Proof of Proposition~\ref{prop:pure}}\label{proof:pure}

Fix $(x,x')\in\mathcal R$. Suppose that some output $y$ belongs to $S(x)$ but not to $S(x')$. By the definition of support,
$M(y\mid x)>0$ and $M(y\mid x')=0$. Pure local privacy applied to the singleton event $\{y\}$ would require
\begin{equation}
M(y\mid x)\leq e^\varepsilon M(y\mid x')=0,
\end{equation}
which is impossible. Thus $S(x)\subseteq S(x')$ for every ordered protected pair. Symmetry of $\mathcal R$ allows the same argument with the inputs reversed and yields $S(x')\subseteq S(x)$.

For a finite-radius line channel, choose inputs separated by more than $2t$. Their radius-$t$ supports are disjoint, so every output in one support is absent from the other. Because ordinary all-pairs privacy protects this pair, the first part rules out any finite privacy parameter.

\subsection{Proof of Proposition~\ref{prop:pair-defect}}\label{proof:pair-defect}

Translation invariance and reflection symmetry reduce the calculation to the ordered pair $(0,h)$ with $h\geq1$. Write
\begin{equation}
S_0=\{-t,\ldots,t\},\qquad S_h=\{h-t,\ldots,h+t\}.
\end{equation}
Using~\eqref{eq:decomposition}, the defect is the mass assigned by the row for input zero to $S_0\setminus S_h$, plus any positive likelihood-ratio excess on $S_0\cap S_h$.

If $h>2t$, the two supports are disjoint. The event $S_0$ has probability one under input zero and probability zero under input $h$, so the ordered defect equals one.

Now suppose $1\leq h\leq t$. The mismatch set is
\begin{equation}
S_0\setminus S_h=\{-t,-t+1,\ldots,h-t-1\},
\end{equation}
which has $h$ elements and does not contain the truthful output zero. Every point in this set has mass $q$, giving support leakage $hq$. On the overlap, output zero has masses $p$ and $q$ in the two rows and contributes $[p-e^\varepsilon q]_+$. Output $h$ contributes $[q-e^\varepsilon p]_+=0$ because $p\geq q$, and every other common output contributes $[q-e^\varepsilon q]_+=0$. The defect is therefore $hq+[p-e^\varepsilon q]_+$.

Finally, let $t<h\leq2t$. The mismatch set still has $h$ elements, but now it contains the truthful output zero. Its total mass is $p+(h-1)q$. Neither truthful output lies in the overlap, so each common output has mass $q$ under both inputs and contributes no positive excess. This gives the final branch.

\subsection{Proof of Corollary~\ref{cor:range-defect}}\label{proof:range-defect}

The required defect is the maximum of~\eqref{eq:pair-defect} over integers $1\leq h\leq H$. If $H>2t$, the protected range contains a separation with disjoint supports. Its defect is one, which is also the largest possible hockey-stick defect between probability distributions.

If $H\leq t$, every protected separation uses the second branch of~\eqref{eq:pair-defect}. The positive-part term is constant in $h$ and $hq$ is nondecreasing, so the maximum is attained at $h=H$.

It remains to consider $t<H\leq2t$. The defect increases with $h$ within each overlap regime. The largest second-regime value is attained at $h=t$, while the largest third-regime value is attained at $h=H$. When $p-e^\varepsilon q\leq0$, the latter value satisfies
\begin{equation}
p+q(H-1)\geq Hq\geq tq.
\end{equation}
When $p-e^\varepsilon q>0$, their difference is
\begin{align}
&p+q(H-1)-\bigl(tq+p-e^\varepsilon q\bigr)\nonumber\\
&\hspace{4em}=q(H-1-t+e^\varepsilon)\geq0.
\end{align}
Thus the endpoint $H$ always determines the range defect. Substituting $p=1-2tq$ into $p+q(H-1)$ gives $1-(2t-H+1)q$.

\subsection{Proof of Theorem~\ref{thm:joint}}\label{proof:joint}

Normalization and truthful bias reduce the fixed-radius parameter set to
\begin{equation}
p=1-2tq,\qquad 0\leq q\leq\frac{1}{2t+1}.
\label{eq:proof-domain}
\end{equation}
The line distortion $qt(t+1)$ is strictly increasing in $q$, so it suffices to identify the smallest point in~\eqref{eq:proof-domain} satisfying the exact privacy constraint.

First let $H\leq t$ and set $a=2t+e^\varepsilon$. Substitution into~\eqref{eq:range-defect} gives
\begin{equation}
\delta_t(q)=Hq+[1-aq]_+
=\begin{cases}
1-(a-H)q,&0\leq q\leq1/a,\\
Hq,&q\geq1/a.
\end{cases}
\label{eq:vshape}
\end{equation}
The first branch decreases and the second increases. Since $e^\varepsilon\geq1$, the kink $1/a$ belongs to~\eqref{eq:proof-domain}. The minimum defect is $H/a$. Hence the radius is feasible exactly when $\delta\geq H/a$. Under this condition, the smallest feasible $q$ lies on the decreasing branch and solves $1-(a-H)q=\delta$, giving the first line of~\eqref{eq:qstar}. The feasibility condition also ensures that this solution does not exceed the kink.

Next let $t<H\leq2t$. The defect becomes
\begin{equation}
\delta_t(q)=1-(2t-H+1)q,
\end{equation}
which decreases throughout~\eqref{eq:proof-domain}. Its minimum is attained at $q=(2t+1)^{-1}$ and equals $H/(2t+1)$. Feasibility and inversion at level $\delta$ give the second line of~\eqref{eq:qstar}. When $H>2t$, the defect equals one and no target $\delta<1$ is feasible.

For every fixed target, feasible radii exist because $H/(2t+e^\varepsilon)$ tends to zero as $t$ tends to infinity. For $t\geq H$, the optimized distortion is
\begin{equation}
D_t^*=\frac{(1-\delta)t(t+1)}{2t+e^\varepsilon-H},
\end{equation}
which diverges linearly with $t$. Consequently, only finitely many radii can improve on any chosen feasible radius, and the minimum over the integer radii is attained.

\subsection{Proof of Proposition~\ref{prop:qinterval}}\label{proof:qinterval}

For $H\leq t$, the two branches in~\eqref{eq:vshape} give separate necessary and sufficient inequalities. On the decreasing branch, $\delta_t(q)\leq\delta$ is equivalent to
\begin{equation}
q\geq\frac{1-\delta}{2t+e^\varepsilon-H}.
\end{equation}
On the increasing branch, it is equivalent to $q\leq\delta/H$. The target is feasible precisely when the two branchwise sets meet at the kink. Their union, intersected with the truthful-bias domain, is the first interval in~\eqref{eq:qinterval}.

For $t<H\leq2t$, direct inversion of the affine defect gives
\begin{equation}
q\geq\frac{1-\delta}{2t-H+1}.
\end{equation}
Intersecting with~\eqref{eq:proof-domain} yields the second interval. The disjoint-support regime and every radius whose minimum defect exceeds the target have empty feasible sets.

\subsection{Proof of Proposition~\ref{prop:sensitivity}}\label{proof:sensitivity}

Inside either feasible branch, $q_t^*$ has the form $(1-\delta)/B$ for a positive denominator $B$ that does not depend on $\delta$. Hence
\begin{equation}
\frac{\partial q_t^*}{\partial\delta}=-\frac{1}{B}<0.
\end{equation}
Line distortion is the positive multiple $t(t+1)q_t^*$, and $p_t^*=1-2tq_t^*$, giving the corresponding directions for distortion and truthful mass.

Within either overlap regime, the denominator decreases with $H$ at unit rate, so $q_t^*$ increases with $H$. For $H\leq t$, differentiation with respect to $\varepsilon$ gives
\begin{equation}
\frac{\partial q_t^*}{\partial\varepsilon}
=-\frac{(1-\delta)e^\varepsilon}
{(2t+e^\varepsilon-H)^2}<0.
\end{equation}
For $t<H\leq2t$, the denominator $2t-H+1$ contains no $\varepsilon$, proving independence in that regime. The claims for $p_t^*$ and distortion again follow from their affine dependence on $q_t^*$.

\subsection{Proof of Proposition~\ref{prop:budget}}\label{proof:budget}

Consider first $H\leq t$. At separation $H$, the mismatch contains $H$ nontruthful outputs of mass $q_t^*$, so
\begin{equation}
L_t^*=Hq_t^*=\frac{H(1-\delta)}{a-H}.
\end{equation}
Feasibility places the optimizer on the decreasing branch of~\eqref{eq:vshape}, where the truthful output contributes the only positive overlap excess. Hence
\begin{align}
E_t^*&=1-aq_t^*
=1-\frac{a(1-\delta)}{a-H}\nonumber\\
&=\frac{a\delta-H}{a-H}.
\end{align}
The numerator is nonnegative exactly because $\delta\geq H/a$. Adding the two terms cancels the common numerator and gives $\delta$.

If $t<H\leq2t$, the truthful output belongs to the mismatch set and every common output has mass $q$ under both rows. Thus the overlap-excess term is zero. The fixed-radius choice in~\eqref{eq:qstar} solves the affine mismatch defect at equality, so its support-leakage term is the full target $\delta$.

\subsection{Proof of Corollary~\ref{cor:floor}}\label{proof:floor}

In the disjoint-support regime, the defect is identically one. In the strong-overlap regime,~\eqref{eq:vshape} is minimized at its kink $q=(2t+e^\varepsilon)^{-1}$, where it equals $H/(2t+e^\varepsilon)$. In the partial-overlap regime, the affine defect decreases over the truthful-bias domain and is minimized at $q=(2t+1)^{-1}$, where it equals $H/(2t+1)$. These minima give all branches of~\eqref{eq:floor}, and feasibility is equivalent to the target exceeding the appropriate minimum.

\subsection{Proof of Proposition~\ref{prop:inverse-budget}}\label{proof:inverse-budget}

At radius $t$, the distortion constraint and truthful-bias domain give
\begin{equation}
 0\leq q\leq \min\left\{\frac{B}{t(t+1)},\frac{1}{2t+1}\right\}.    \label{eq:proof-budget-domain}
\end{equation}
If $H>2t$, a protected pair has disjoint supports and the defect is one for every admissible $q$, which proves the first branch.

Suppose $H\leq t$ and write $a=2t+e^\varepsilon$. Equation~\eqref{eq:vshape} decreases on $[0,1/a]$ and increases after $1/a$. Since $1/a<1/(2t+1)$, minimizing it over~\eqref{eq:proof-budget-domain} gives
\begin{equation}
 \widehat q_t(B)=\min\left\{\frac{B}{t(t+1)},\frac1a\right\}.       \label{eq:proof-budget-q-strong}
\end{equation}
If the first term is active, substitution into the decreasing branch gives
$1-(a-H)B/[t(t+1)]$. If the kink is active, the defect is the floor $H/a$. Taking the larger of these two affine pieces gives the strong-overlap branch of~\eqref{eq:inverse-budget}.

Now suppose $t<H\leq2t$. The defect
$1-(2t-H+1)q$ decreases over the entire truthful-bias domain. Its optimizer under~\eqref{eq:proof-budget-domain} is therefore
\begin{equation}
 \widehat q_t(B)=\min\left\{\frac{B}{t(t+1)},\frac1{2t+1}\right\}.  \label{eq:proof-budget-q-partial}
\end{equation}
Substitution gives either the budget-limited affine term or the floor $H/(2t+1)$, proving the remaining branch.

It remains to verify attainment over radii. When $B=0$, every radius has defect one. Let $B>0$. For any sufficiently large $t\geq H$, the first term in~\eqref{eq:proof-budget-q-strong} is active and positive, so $\underline\delta_t(B)<1$. On the other hand,
\begin{equation}
 \lim_{t\to\infty}\left[1-
 \frac{(2t+e^\varepsilon-H)B}{t(t+1)}\right]=1,       \label{eq:proof-budget-tail}
\end{equation}
and the floor tends to zero. Hence $\underline\delta_t(B)\to1$. Choose a finite radius with value strictly below one. The tail is eventually worse than that value, leaving only finitely many candidate radii that can attain the infimum. The global minimum is therefore attained.

\subsection{Proof of Corollary~\ref{cor:minradius}}\label{proof:minradius}

Radii below $\lceil H/2\rceil$ have disjoint supports for at least one protected separation and therefore have floor one. On the integer interval $\lceil H/2\rceil\leq t<H$, the floor is $H/(2t+1)$ and decreases with $t$. For $t\geq H$, it is $H/(2t+e^\varepsilon)$ and again decreases. At the transition,
\begin{equation}
\frac{H}{2H+e^\varepsilon}<\frac{H}{2H-1},
\end{equation}
so the full sequence of floors is nonincreasing.

If $\delta\geq H/(2H-1)$, radius $H-1$ is feasible whenever that transition regime exists. The smallest feasible radius is therefore found in the partial-overlap branch by solving
\begin{equation}
\frac{H}{2t+1}\leq\delta,
\end{equation}
together with $t\geq\lceil H/2\rceil$. Integer rounding gives the first branch of~\eqref{eq:minradius-all}. If $\delta<H/(2H-1)$, no radius below $H$ is feasible. Solving $H/(2t+e^\varepsilon)\leq\delta$ subject to $t\geq H$ gives the second branch. For $H=1$, the threshold equals one, so every allowed $\delta$ correctly uses the second branch.

\subsection{Proof of Corollary~\ref{cor:support}}\label{proof:support}

Suppose first that $t<H$. If $H>2t$, the radius is infeasible because its privacy floor is one. Otherwise $t<H\leq2t$, and
\begin{equation}
\delta_t^{\min}=\frac{H}{2t+1}\geq\frac{H}{2H-1}>\delta.
\end{equation}
Thus the stated condition forces every feasible radius to satisfy $t\geq H$. In this regime feasibility requires
\begin{equation}
\delta\geq\frac{H}{e^\varepsilon+2t}.
\end{equation}
Rearrangement and $s=2t+1$ yield~\eqref{eq:support-bound}. The order statement follows with fixed $H$ and $\varepsilon$.

\subsection{Proof of Corollary~\ref{cor:candidates}}\label{proof:candidates}

Every $t<\lceil H/2\rceil$ has $H>2t$ and is infeasible for $\delta<1$. The only candidates below $H$ therefore lie in the finite transition range specified by $\mathcal T_H$, and Corollary~\ref{cor:floor} supplies their exact feasibility test.

For $t\geq H$, let $c=e^\varepsilon-H$ and ignore the positive constant $1-\delta$. The optimized distortion is proportional to
\begin{equation}
f(t)=\frac{t(t+1)}{2t+c}.
\end{equation}
For real $t\geq H$, its derivative has the sign of
\begin{equation}
N(t)=2t^2+2ct+c.
\end{equation}
At the left endpoint,
\begin{equation}
N(H)=e^\varepsilon(2H+1)-H>0.
\end{equation}
Moreover, $N'(t)=4t+2c\geq2H+2e^\varepsilon>0$ on this domain. Hence $f$ is strictly increasing for $t\geq H$. The first feasible integer in this regime is exactly $t_0$, and no larger radius can be optimal. Counting the transition integers plus $t_0$ gives at most $\lfloor H/2\rfloor+1$ candidates.

\subsection{Proof of Corollary~\ref{cor:cost}}\label{proof:cost}

For each displacement $j\in\{1,\ldots,t\}$, the reports $x-j$ and $x+j$ both occur with probability $q$ and both incur cost $g(j)$. The truthful report has zero cost. Summation therefore gives~\eqref{eq:generalcost}. Its coefficient of $q$ is strictly positive under the stated assumption, so the cost is minimized by the same smallest privacy-feasible $q$ as line distortion. Substituting $g(j)=j^r$ gives the power-cost expression.

\subsection{Proof of Corollary~\ref{cor:ratio}}\label{proof:ratio}

At fixed $t$, both line distortions contain the common positive factor $t(t+1)$. Their ratio is therefore $q_t^*/q_{\mathrm{LR}}$. Substitution of $q_t^*$ and $q_{\mathrm{LR}}$ gives~\eqref{eq:ratio}. The ratio is at most one precisely when
\begin{equation}
(1-\delta)(2t+e^\varepsilon)\leq2t+e^\varepsilon-H,
\end{equation}
which rearranges to $\delta\geq H/(2t+e^\varepsilon)$. Equality holds at the privacy floor and only there.

\subsection{Proof of Corollary~\ref{cor:asymptotics}}\label{proof:asymptotics}

For sufficiently small $\delta$, Corollary~\ref{cor:support} excludes every radius below $H$, and Corollary~\ref{cor:candidates} makes $t^*=t_0$. There is a bounded remainder $r_\delta\in[0,2)$ such that
\begin{equation}
2t^*=\frac{H}{\delta}-e^\varepsilon+r_\delta.
\label{eq:rounding-remainder}
\end{equation}
The first two expansions in~\eqref{eq:asymptotics} follow immediately.

Using~\eqref{eq:rounding-remainder} in the strong-overlap branch of~\eqref{eq:qstar} gives
\begin{align}
q^*&=\frac{1-\delta}{H/\delta-H+r_\delta}\nonumber\\
&=\frac{\delta}{H}
\frac{1-\delta}{1-\delta+(r_\delta/H)\delta}
=\frac{\delta}{H}+O(\delta^2).
\end{align}
Normalization can be written in a form that avoids cancellation:
\begin{align}
p^*&=\frac{e^\varepsilon-H+2t^*\delta}
{2t^*+e^\varepsilon-H}\nonumber\\
&=\frac{e^\varepsilon+(r_\delta-e^\varepsilon)\delta}
{H/\delta-H+r_\delta}
=\frac{e^\varepsilon\delta}{H}+O(\delta^2).
\end{align}
Finally, $t^*=H/(2\delta)+O(1)$ and $q^*=\delta/H+O(\delta^2)$ imply
\begin{equation}
q^*t^*(t^*+1)=\frac{H}{4\delta}+O(1),
\end{equation}
which completes the proof.

\end{document}